\documentclass[showpacs,reprint,nofootinbib,longbibliography,superscriptaddress,prd]{revtex4-1}

\usepackage{hyperref}
\usepackage[utf8x]{inputenc}
\usepackage{epsfig,dsfont}
\usepackage{color}
\usepackage{graphicx}
\usepackage[normalem]{ulem}
\usepackage{color}
\usepackage{verbatim}

\usepackage{amsmath}
\usepackage{amsfonts}
\usepackage{amssymb}
\usepackage{cases}

\newcommand{\la}{\mathcal{L}}

\def\twis{t}

\newcommand{\et}{\eta}
\newcommand{\ett}{\eta_{\text{\,\twis}}}

\renewcommand{\th}{\theta}
\newcommand{\tht}{\theta_{\text{\twis}}}
\newcommand{\ph}{\phi}
\newcommand{\pht}{\phi_{\text{\twis}}}

\renewcommand{\u}{u}
\newcommand{\ut}{u_{\text{\twis}}}
\renewcommand{\v}{v}
\newcommand{\vt}{v_{\text{\twis}}}

\newcommand{\rt}{r_{\text{\twis}}}
\newcommand{\st}{s_{\text{\twis}}}

\newcommand{\n}{n}
\newcommand{\nt}{n_{\text{\twis}}}

\newcommand{\mt}{m_{\text{\twis}}}

\newcommand{\vv}{v}
\newcommand{\vvt}{v_{\text{\twis}}}

\newcommand{\wt}{w_{\text{\twis}}}

\newcommand{\pbar}[1]{\overset{\scriptscriptstyle{\,(-)}}{#1}}
\newcommand{\M}{M}
\newcommand{\C}{\mathbb{C}}

\newcommand{\re}{\text{Re}\,}
\newcommand{\im}{\text{Im}\,}
\newcommand{\ve}[2]{\left(\begin{smallmatrix}#1\\#2\end{smallmatrix}\right)}

\newcommand{\DD}{\mathcal{D}}

\newcommand{\para}{|}

\usepackage{tikz}
\usetikzlibrary{arrows,calc,shapes,decorations.pathreplacing}

\begin{document}

\title{Complex instantons in sigma models with chemical potential}

\author{Falk Bruckmann}
\affiliation{Universit\"at Regensburg, Institut f\"ur Physik, Universit\"atsstra{\ss}e 31, 93053 Regensburg, Germany}

\author{Stephan Lochner}
\affiliation{Universit\"at Regensburg, Institut f\"ur Physik, Universit\"atsstra{\ss}e 31, 93053 Regensburg, Germany}

\begin{abstract}
 We analyze two-dimensional nonlinear sigma models at nonzero chemical potentials, which are governed by a complex action. In the spirit of contour deformations (thimbles) we extend the fields into the complex plane, which allows to incorporate the chemical potentials $\mu$ as twisted boundary conditions. We write down the equations of motion and find exact BPS-like solutions in terms of pairs of (anti)holomorphic functions, in particular generalizations of unit charge and fractional instantons to generic $\mu$. The decay of these solutions is controled by the imaginary part of $\mu$ and a vanishing imaginary part causes jumps in the action.
 We analyze how the total charge is distributed into localized objects and to what extent these are characterized by topology.
\end{abstract}


\maketitle

\section{Introduction}

Two-dimensional nonlinear sigma models have been known for a long time to share nontrivial properties -- such as asymptotic freedom, dynamical mass generation, topology, supersymmetric extensions etc.\ -- with four-dimensional nonabelian gauge theories, see e.g.\ \cite{Shifman:2012zz}. More recent research has focussed on refined similarities between these models: instanton constituents on compactified space-times \cite{Eto:2004rz,Eto:2006mz,
Bruckmann:2007zh,Brendel:2009mp,Harland:2009mf}, with which the asymptotic perturbation theory may be tamed (trans-series/renormalons) \cite{Dunne:2012zk}, 't Hooft anomalies \cite{Gaiotto:2017yup,Tanizaki:2017qhf,Sulejmanpasic:2018upi} and the sign problem at nonzero density.

Indeed, we will analyze (purely bosonic) sigma models at \textit{nonzero chemical potentials} $\mu$. Since the action becomes complex, numerical simulations are hampered severly. Purely imaginary $\mu$'s do not cause such a problem and will be an exceptional case in most of what we discuss. At real $\mu$ (only) dualizing the lattice theory has been shown to solve the sign problem \cite{Bruckmann:2015sua}. When the chemical potential equals the mass gap, the system undergoes a quantum phase transition 
\cite{Bruckmann:2016txt}. 

When treating these systems by stochastic quantization (Langevin dynamics; in the original field representation), the drift from the complex action immediately drives the field configurations into the complex plane. The thimble method\footnote{The thimble and Langevin method not only share the same saddles, but were empirically found to concentrate around similar complex configurations \cite{Aarts:2014nxa}.} relies on field manifolds with constant phase of the (path) integral weights, which cannot be realized on real fields either. Complexifying the fields has actually a long history in sigma models: The large $N$ proof of the sigma model mass gap relies on a complex saddle of the Lagrange multiplier \cite{D'Adda:1978kp,Witten:1978bc}. 

Our work is motivated by the continuum trans-series description, for which  sigma models serve as a showcase \cite{Dunne:2012zk,Dunne:2012ae,Misumi:2014jua,Dunne:2015ywa}. This approach requires the knowledge of classical solutions -- typically with topological features -- that could possibly be combined into neutral molecules to contribute to the vacuum sector etc. As we have argued, for generic $\mu$'s it is natural to \textit{seek classical solutions in the complexified fields}. Note that the powerful Bogomolny-Prasad-Sommerfield (BPS) bound \cite{Bogomolny:1975de,Prasad:1975kr}, that reduces the order of the differential equations which are then easily solved by (anti)holomorphic fields, relies on a completed square and loses its power for complex numbers. We will nonetheless be able to solve the equation of motion, after having pushed the chemical potential into the boundary conditions. The latter equivalence has been utilized in sigma models with twisted boundary conditions \cite{Eto:2004rz,Eto:2006mz,
Bruckmann:2007zh,Brendel:2009mp,Harland:2009mf}, which could be seen as purely imaginary chemical potential applied in the spatial direction\footnote{For bosons twisted boundary conditions in space and time are interchangeable (when exchanging length and inverse temperature), only fermions could distinguish between space and time.}. 

The objects presented here are to our knowledge the first complex solutions of a field theory complexified in this way. In the sine-Gordon-like quantum mechanics that results from dimensionally reducing (supersymmetric) sigma models (at $\mu=0$), complex saddles and bions have been discussed in \cite{Fujimori:2016ljw}. We focus on the derivation and description of complex BPS-like solutions at complex $\mu$. Many of the nice properties of sigma models are modified during the complexification of the fields, therefore, we will repeat them in detail to emphasize which are crucially changed in the complex setting. For instance, the most useful representations of these models (with respect to solutions) are already complex, such that the `complexification' of these representations has to be done with care. 

The obtained solutions consist again of (anti)holomorphic functions, however, we have to specify twice as many functions for the doubled degrees of freedom. These objects are (anti)selfdual in the sense that their action density equals (minus) the topological density. The latter is still a total derivative and given in terms of the Laplacian of a logarithm, but the meaning of the complex total topological charge as a winding number is more intricate, as are the symmetries of the complexified system.

As examples we analyze analogues of fractional constituents and unit charge instantons in the O(3)$\cong$CP(1) model at generic $\mu$. Their densities turn out to be analytic continuations of the corresponding densities from imaginary $\mu$ (i.e.\ from twisted solutions) to generic $\mu$. This has important consequences for the decay of these densities and thus the total actions/charges of these objects. The fractional instantons, for instance, have finite action only when an imaginary part of $\mu$ is present. From the viewpoint of real $\mu$, therefore, an imaginary $\mu$ might be viewed as a `regulator'. The limit of vanishing imaginary part of $\mu$, however, produces an imaginary jump in the total action/charge, similar to lateral Borel resummations of sign-coherent series. Technically, these jumps appear at branch cuts of the square root function. This also holds for the analogue of the unit charge instantons, where the imaginary part of the total action/charge jumps at vanishing imaginary part of $\mu$ and a complementary $\bar{\mu}$, between which the topological charge is unity.

This work is organized into two main parts, one about the O(3) model and its specific realizations and one about the more general CP(N-1) models. In both parts, we first discuss (conventional) BPS solutions, the global symmetries to which chemical potentials couple, the method of pushing the latter into twisted boundary conditions and the resulting complexity issue. Then we perform the field complexification doubling the degrees of freedom and write down the complex field equations and their solutions in general. At the end of each part we discuss basic examples, fractional constituents and unit charge instantons. Sec.~\ref{sec:summary} contains our summary and outlook.

\section{O(3) model}

There are (at least) three ways to parametrize the O(3) field, and for reasons of illustration we will discuss them in parallel. The defining parametrization uses a real three-vector field $\vec{\et}(x)$ normalized to 1, $\vec{\et}^{\,2}(x)=1$. In the following we will omit arguments $x$ and the vector arrow on $\eta$. Polar angle fields $\th\in[0,\pi]$, $\ph\in[0,2\pi]$ can be introduced via $\et_1=\sin\th\cos\ph$ etc.\ and a  complex stereographic field via 
\begin{align}
 \u
 =\frac{\et_1+i\et_2}{1-\et_3}
 = \cot\frac{\th}{2}\,e^{i\ph}\,.
 \label{eq_o3_field_trafo}
\end{align}
The Langrangian without chemical potential reads
 \begin{subnumcases}
 {2\mathcal{L}_0=\label{eq:lagrange_orig}}
     (\partial_\nu\et)^2\,, 
     \label{eq:lagrange_orig1}\\
     (\partial_\nu\th)^2+\sin^2\!\th\,(\partial_\nu\ph)^2\,,\label{eq:lagrange_orig2}\\
     4\,\dfrac{\partial_\nu \u\, \partial_\nu u^*}{(1+|u|^2)^2}\,.
     \label{eq:lagrange_orig3}
  \end{subnumcases}
The interesting features of this system are caused by its nonlinearity, which is manifest in the latter two paramterizations, in the $\et$-parametrization it is caused by the constraint.

\subsection{Preparation: BPS solutions, symmetries, chemical potential and twisted fields}
\label{sec:all_till_bps_solns}

A famous tool to obtain classical solutions is the Bogomolnyi bound, in which the Lagrangian is split into an (absolute) square plus (or minus) a topological term. This is most transparent in the $\u$-picture:
\begin{align}
 \!\!\!\!\la_0(\u)
 &=4\,\frac{|\partial\u|^2+|\partial^*\!\u|^2}{(1+|\u|^2)^2}\\
 &=8\left(\frac{|\partial^*\!\u|}{1+|\u|^2}\right)^2\!\!+4\pi q
 =8\left(\frac{|\partial\u|}{1+|\u|^2}\right)^2\!\!-4\pi q\,,
 \label{eq:o3_bound}
\end{align}
where we introduced complex coordinates and derivatives
\begin{align}
 z^{(*)}=x_1\pm i x_2\,,\qquad 
 \partial^{(*)}=(\partial_1\mp i\partial_2)/2\,,
 \label{eq:complex_coord}
\end{align}
and the topological charge density
\begin{align}
 q
 &=\frac{1}{\pi}\frac{|\partial\u|^2-|\partial^*\!\u|^2}{(1+|\u|^2)^2}\,,
\end{align}
which is a total derivative (see, e.g., Eqs.~\eqref{eq:q_tot_der} and \eqref{eq:q_tot_der_A} below with $v=u$).

Consequently, configurations with $\la_0=\pm 4\pi q$ possess minimal action in a given topological sector and thus also solve the equations of motion. For these BPS solutions, the argument of the square must vanish, which means a first order differential equation that can immediately be solved: $\u=\u(z^{(*)})$ are \textit{(anti)meromorphic functions for positive (negative) topological charge.} Poles in $\u$ are admissible as they represent the north pole $\et=(0,0,1)^T$ (zeros of $u$ represent the south pole).

For BPS solutions, a compact formula for the topological charge density applies, $q=\pm1/4\pi\cdot\Delta \log(1+|u|^2)$
with the Laplacian $\Delta=4\,\partial\partial^*$.

\bigskip

From now on we will consider a nonzero temperature $T=1/\beta$ represented by periodic boundary conditions of the fields under $x_2\to x_2+\beta$. Compatible BPS solutions are naturally analyzed through a Fourier expansion \cite{Brendel:2009mp} in the (anti)holomorphic coordinate $2\pi /\beta \cdot z^{(*)}$. 

One of the three global O(3) symmetry-rotations shifts $\ph\to \ph+\alpha$ and manifests itself as O(2)-rotations of the first two components of $\vec{\et}$ and U(1)-multiplications $\u\to\u\exp(i\alpha)$, respectively. Coupling a chemical potential to the corresponding conserved charge, the Lagrangian changes according to:
\begin{subnumcases}
 {\!\!\!\!\!\!\!\!\!\!\!\!\!\mathcal{L}_\mu\!=\!\mathcal{L}_0\text{ w/}\,
 \label{eq:newlag}}
     \partial_2\et\!\to\!(\partial_2-i\mu\, T^3)\et\,, \label{eq:newlag_1}\\
     \partial_2\ph\!\to\!\partial_2\ph+i\mu\,, \label{eq:newlag_2}\\
     \partial_2 \u\!\to\! (\partial_2-\mu)\u\,,\: 
     \partial_2 \u^*\!\to\! (\partial_2+\mu)\u^*, \label{eq:newlag_3}
  \end{subnumcases}
where $(T^3)_{ab}=\epsilon_{3ab}$ (is antisymmetric). For real $\mu$, the time derivative of $\ph$ receives an imaginary part, while in the other two versions $\partial_2$ receives a hermitian part, which makes the action complex. 

In all the three cases one could try to revert the new Lagrangian $\mathcal{L}_\mu$ to the original $\mathcal{L}_0$ without chemical potential by redefining the fields compensating the modifications in Eq.~\eqref{eq:newlag}, e.g., defining $\partial_2\ph+i\mu=:\partial_2\pht$. To be precise, we define `twisted fields'
\begin{subequations}\begin{align}
 \ett
 &:= e^{-i\mu T^3 x_2} \et\,, 
 \label{eq:def_twisted1}\\
 \pht
 &:=i\mu x_2+\ph\,, \qquad \tht=\th\,,
 \label{eq:def_twisted2}\\
 \ut 
 &:=e^{-\mu x_2}\u\,,\qquad\,
 \ut^*:=e^{\,\mu x_2}\u^*\qquad(\text{imag.}\,\mu)\,,
 \label{eq:def_twisted3} 
\end{align}
\label{eq:def_twisted}\end{subequations}
such that the Lagrangian indeed obeys
\begin{align}
 \la_\mu(\Phi)=\la_{\mu=0}(\Phi_{\text{\twis}})\,,\quad
 \Phi\in\{\et,(\th,\ph),\u\}\,.
 \label{eq:relating_lagrangians}
\end{align}
This relation means in particular that if the periodic fields $\Phi$ extremize $\mathcal{L}_\mu$, then the twisted fields $\Phi_{\text{\twis}}$ extremize $\mathcal{L}_{\mu=0}$ and vice versa. Thus, properly periodic solutions at nonzero $\mu$ can be obtained from twisted solutions at $\mu=0$. These fields are called twisted since they obey,
\begin{subequations}
 \begin{align}
 \ett(x_2+\beta)
  &=e^{-i\mu T^3 \beta}\ett(x_2)\,,
  \label{eq:twisted_bc1}\\
 \pht (x_2+\beta)
  &=i\mu\beta +\pht(x_2)\,,\:\:
 \tht(x_2+\beta)=\tht(x_2)\,,
 \label{eq:twisted_bc2}\\
 \ut^{(*)}(x_2+\beta)
  &=e^{\mp\mu\beta}\ut^{(*)}(x_2)\qquad(\text{imag.}\,\mu)\,.
  \label{eq:twisted_bc3}
 \end{align}
 \label{eq:twisted_bc}%
\end{subequations}
and we seem to have absorbed the chemical potential completely in these boundary conditions. 

For generic $\mu$, however, the definitions of twisted fields and boundary conditions \textit{conflict with the nature of the fields}: the rotation in Eq.~\eqref{eq:def_twisted1} does not belong to O(2), the additional term in Eq.~\eqref{eq:def_twisted2} is not real and Eqs.~\eqref{eq:def_twisted3} are not compatible with complex conjugation. While for purely imaginary $\mu$, these problems do not occur and this method has been used to write down classical solutions (see below), for generic $\mu$ we will first complexify the fields properly, see Sec.~\ref{sec:o3_complexif}\footnote{As will be shown there, Eqs.~\eqref{eq:def_twisted1}, \eqref{eq:def_twisted2}, \eqref{eq:twisted_bc1} and \eqref{eq:twisted_bc2} will remain valid at generic $\mu$ \textit{with complex fields}, where as Eqs.~\eqref{eq:def_twisted3} and \eqref{eq:twisted_bc3} will be replaced by Eqs.~\eqref{eq_complex_conjugate} and \eqref{eq_bc_o3_complex}.}.

\bigskip

With the help of twisted fields, BPS solutions can be extended to the case of \textit{purely imaginary chemical potential}, say
\begin{align}
 \mu=-2\pi i\omega/\beta\,,\quad 
 \omega\in[0,1]\,,
 \label{eq:mu_omega}
\end{align}
where $\omega$ is a real twist \cite{Eto:2004rz,Eto:2006mz,
Bruckmann:2007zh,Brendel:2009mp,Harland:2009mf}, since the twisted boundary condition \eqref{eq:twisted_bc3} is now compatible with complex conjugation. This boundary condition can easily be satisfied by an additional (anti)holomorphic factor $\exp(2\pi  \omega z^{(*)}/\beta)$ multiplying a (anti)holomorphic periodic function. 
In this setting the total topological charge can be integer -- for `full instantons' -- or fractional with fractional part $\pm\,\omega$ -- when including `instanton constituents'. The analogy to these solutions will help us to solve the case of general $\mu$ below.

\subsection{Complexification}
\label{sec:o3_complexif}

We have already argued that the action \eqref{eq:newlag} is complex for generic chemical potentials. The boundary conditions \eqref{eq:twisted_bc} for the first two parametrizations suggest to use \textit{complexified fields} $(\et_1,\et_2)_{\text{\twis}}$ and $\pht$ for solutions of the equations of motion. We will also promote $\et_{3,\text{\twis}}$ and $\tht$ to complex fields, because they couple to the former. In this way we have doubled the two real degrees of freedom in the angle representation.

Keeping the equivalence of the three parametrizations, e.g., $\et_{\text{\twis},1}=\sin\tht\cos\pht$ etc.\ means that $\ett$ becomes a complex vector with\footnote{To obtain \eqref{eq:o3_new_constraint}, one uses relations like $\sin^2(\ldots)+\cos^2(\ldots)=1$ etc., which remain valid for complex arguments.}
\begin{align}
 \ett^2=1\,,\qquad \ett\in\mathbb{C}^3\,.
 \label{eq:o3_new_constraint}
\end{align}
This relation does not contain a complex conjugate on the left hand side, so it represents two real constraints for three complex fields. Thus also the number of real degrees of freedom in $\ett$ has doubled from two to four.

These complexified fields are required to obey the twisted boundary consitions \eqref{eq:twisted_bc1} and \eqref{eq:twisted_bc2}. We will use the same functional forms of the Lagrangian, \eqref{eq:newlag_1} and  \eqref{eq:newlag_2}, in terms of these complex fields which, therefore, is a holomorphic function of them (in other words, the complex conjugate fields do not appear) and again, Eq.~\eqref{eq:relating_lagrangians} will relate solutions of these complex twisted fields to complex solutions at nonzero $\mu$. 

\bigskip

Since the stereographic coordinate $\ut$ is complex from the beginning, the `complexification' of it needs to be done carefully. One way to achieve it is to complexify the real degrees of freedom contained in it, see also \cite{Fujimori:2016ljw}. For that write $\ut=\rt+i\st$
with $\rt=\cot\frac{\tht}{2}\,\cos\pht$ and $\st=\cot\frac{\tht}{2}\,\sin\pht$ (see \eqref{eq_o3_field_trafo}) and insert complexified angles on the right hand sides, which automatically complexify the formerly real variables $(\rt,\st)$ on the left hand sides. 

An equivalent\footnote{The transformation between the two complexification is linear, namely $(\ut,\vt^*)=(\rt+i\st,\rt-i\st)$.} complexification is to treat $\ut$ and its complex conjugate as independent, renaming $\ut^*\to\vt^*$: 
\begin{subequations}
 \begin{align}
 \ut
 &=\cot\frac{\tht}{2}\,e^{i\pht}
 =\frac{\et_{\text{\twis},1}+i\et_{\text{\twis},2}}
 {1-\et_{\text{\twis},3}}\,,\\
 \vt^*
 &=\cot\frac{\tht}{2}\,e^{-i\pht}
 =\frac{\et_{\text{\twis},1}-i\et_{\text{\twis},2}}
 {1-\et_{\text{\twis},3}}\,,
 \label{eq_complex_conjugate_2}
 \end{align}
\label{eq_complex_conjugate}%
\end{subequations}
the reverse transformation is,
\begin{align}
 \tht=2\,\text{arccot}\sqrt{\ut \vt^*}\,,\qquad
 \pht=-\,\frac{i}{2}\,\log \frac{\ut}{\vt^*}\,.
 \label{eq_complex_conjugate_reverse}
\end{align}
From these representations one can derive the corresponding twisted boundary conditions for $\ut$ and $\vt^*$,
\begin{subequations}
 \begin{align}
 \ut(x_2+\beta)
  &=e^{-\mu\beta}\ut(x_2)\,,
  \label{eq_bc_o3_complex_1}\\  
  \vt^*(x_2+\beta)
  &=e^{\,\mu\beta}\vt^*(x_2)\,,
 \end{align}
\label{eq_bc_o3_complex}%
\end{subequations}
as well as their Lagrangian
\begin{align}
 \mathcal{L}_0(\ut,\vt^*)
 =4\,\frac{\partial_\nu \ut\,\partial_\nu \vt^*}{(1+\ut\vt^*)^2}\,.
\end{align}
Note that for generic chemical potential, $\pht$ is complex and $\vt\neq \ut$; for purely imaginary chemical potential, $\pht$ stays real and $\vt=\ut$ (which is why we defined $\vt^*$ in Eq.~\eqref{eq_complex_conjugate_2}). 

Again, due to the analogue of Eq.~\eqref{eq:relating_lagrangians}, 
$\mathcal{L}_\mu(\u,\v^*)=\mathcal{L}_{0}(\ut,\vt^*)$,
the untwisted fields $\u=e^{\,\mu x_2}\ut$, $\v^*=e^{\,-\mu x_2}\vt^*$ are periodic and (complex) solutions at nonzero~$\mu$.

\subsection{Complex solutions}

Complex coordinates and derivatives will again be useful to write the Lagrangian with complexified fields as
\begin{align}
 \mathcal{L}_0(\ut,\vt^*)
 =4\,\frac{\partial\ut\,\partial^*\vt^*+\partial^*\ut\,\partial\vt^*}{(1+\ut\vt^*)^2}\,.
 \label{eq_neq_lang_def}
\end{align}
The equations of motion can be derived from it straightforwardly,
\begin{subequations}
\begin{align}
 \partial\partial^*\ut
 &=\frac{2\vt^*}{1+\ut\vt^*}\,\partial\ut\partial^*\ut\,,\\
 \partial\partial^*\vt^*
 &=\frac{2\ut}{1+\ut\vt^*}\,\partial\vt^*\partial^*\vt^*\,.
\end{align}
\end{subequations}
Although still second order differential equations, they can in fact easily be solved in a BPS-like manner: 
\begin{subequations}\begin{align}
 \ut=\ut(z)\,, \quad\:\, 
 \vt^*&=\vt^*(z^*)
 \qquad (\mathcal{L}_0=+4\pi q)\,,
 \label{eq_the_solutions_o3_1}\\
 \ut=\ut(z^*)\,, \quad 
 \vt^*&=\vt^*(z)
 \qquad\:\: (\mathcal{L}_0=-4\pi q)\,,
 \end{align}
 \label{eq_the_solutions_o3_first}%
\end{subequations} 
or 
\begin{subequations}
 \begin{align}
 \ut&=\ut(z)\,,\quad\, 
 \vt^*=\vt^*(z) \qquad \:\:\,
 (\mathcal{L}_0=0=q)\,,\\
 \ut&=\ut(z^*)\,, \quad 
 \vt^*=\vt^*(z^*)\qquad 
 (\mathcal{L}_0=0=q)\,. 
 \end{align}
\label{eq_the_solutions_o3_second}%
\end{subequations}
In the limit of purely imaginary $\mu$ including $\mu\to 0$, where we have argued that real solutions $\vt=\ut$ exist, the first set of solutions \eqref{eq_the_solutions_o3_first} becomes a single holomorphic/antiholomorphic BPS solution, respectively (since $\vt=\vt(z)$ and $\vt=\vt(z^*)$, respectively). The second set of solutions \eqref{eq_the_solutions_o3_second} can only become constants in this limit. Consistently, the solutions \eqref{eq_the_solutions_o3_second} possess vanishing action. 

It is therefore tempting to see what became of the topological properties for these complex solutions. The formal equivalent of the topological density,
\begin{align}
 q
 &=\frac{1}{\pi}\frac{\partial\ut\,\partial^*\!\vt^*
 -\partial^*\!\ut\,\partial\vt^*}{(1+\ut\vt^*)^2}\,,
\end{align}
appears in the analogue of the BPS relation,
\begin{align}
 \mathcal{L}_0
 =8\frac{\partial^*\!\ut\,\partial\vt^*}{(1+\ut\vt^*)^2}+4\pi q
 =8\frac{\partial\ut\,\partial^*\!\vt^*}{(1+\ut\vt^*)^2}-4\pi q
\end{align}
The first set of solutions \eqref{eq_the_solutions_o3_first} has $\mathcal{L}_0=\pm 4\pi q$ (as in their $\mu\to 0$ BPS limit), whereas for the second set of solutions not only the Lagrangian vanishes but also the topological density.

The topological density $q$ is still a total derivative:
\begin{align}
  q = \frac{i}{2\pi}(\bar{\partial}A - \partial \bar{A})\,,
  \label{eq:q_tot_der}
\end{align}
where $\bar{\partial}= \partial^*$ and
\begin{align}
 \pbar{A} = i \frac{\pbar{\partial}\log(\ut/\vt^*)}{1+\ut\vt^*}\,.
\label{eq:q_tot_der_A}
\end{align}
We postpone the meaning of topology in the complex setting to the corresponding discussion in the CP(N-1) framework, Sec.~\ref{sec:topol}.

\subsection{Remark on O(N) models}

The (real) O(3) model can be trivially embedded into higher O(N) models, through $\vec{\xi} = (0,\hdots,\eta_1,\eta_2,\eta_3,\hdots,0)^T \in \mathbb{R}^N$, $|\vec{\xi}|=1$, which preserves both the global symmetry -- including associated twist and chemical potential -- and the (2nd order) equations of motion for the non-vanishing components. Embedding O(3) BPS solutions, one therefore still obtains classical solutions for O(N). These so-called called unitons \cite{Dunne:2015ywa} are no longer protected by topology (since the relevant homotopy group $\pi_2(S^{N-1}$) is trivial for $N>3$). For our complexification of O(3) instantons, however, this topological protection was not essential, and the procedure should straightforwardly carry over to O(N) models at nonzero chemical potential, yielding `complexified unitons' as the corresponding solutions.

\subsection{Example: constituent}
\label{sec:example_o3}

As an example we consider the simplest solutions of Eq.~\eqref{eq_the_solutions_o3_1}
\begin{align}
 \ut=e^{\,i\mu (z-z^{(1)})}\,,\qquad 
 \vt^*=e^{\,i\mu (z-z^{(2)})^*}\,,
\label{eq:o3_example}
\end{align}
that obey the boundary conditions \eqref{eq_bc_o3_complex}, since $\ut\sim\exp(-\mu x_2)$, $\vt^*\sim\exp(\mu x_2)$.
As we will see, these are the analogues of instanton constituents at purely imaginary $\mu$ discussed around Eq.~\eqref{eq:mu_omega}. 

The product
\begin{align}
 \ut\vt^*=e^{2i\mu (x_1-\Delta x_1)}\,,\qquad
 \Delta x_1=\frac{z^{(1)}+z^{(2)\,*}}{2}
\end{align}
is static and so is $\tht$ according to \eqref{eq_complex_conjugate_reverse}. In this way it obeys its periodic boundary condition \eqref{eq:twisted_bc2}, nonetheless, $\tht$ is complex (unless $\mu$ is purely imaginary). The second angle $\pht$ is proportional to the logarithm of the ratio $\ut/\vt^*$, see Eq.~\eqref{eq_complex_conjugate_reverse}, and thus $x_1$-independent, $\pht=i\mu(x_2-[z^{(1)}-z^{(2)\,*}]/2i)$. This linear dependence on $x_2$ is the simplest way of picking up a factor $i\mu\beta$ in the boundary condition \eqref{eq:twisted_bc2} and $\pht$ is generically complex, as expected.

For the action and topological density we use that $\partial\ut\,\partial^*\!\vt^*=-\mu^2\ut\vt^*$ is static,
too, as is
\begin{align}
 \mathcal{L}_0
 =-\frac{\mu^2}{\cos^2(\mu (x_1+\Delta x_1) )}
 =4\pi q\,,
 \label{eq:constituent_act_dens}
\end{align}
which agrees with the analytic continuation of the corresponding formula in the twisted case \cite{Bruckmann:2007zh} up to the fact that the shift $\Delta x_1$ can also take complex values. However, to achieve $\ut=\vt$ in the limit of purely imaginary $\mu$, the parameters must fulfil $z^{(1)}=z^{(2)}$, and $\Delta x_1$ is real. 

\begin{figure}[t]
\protect \includegraphics[width=\linewidth]{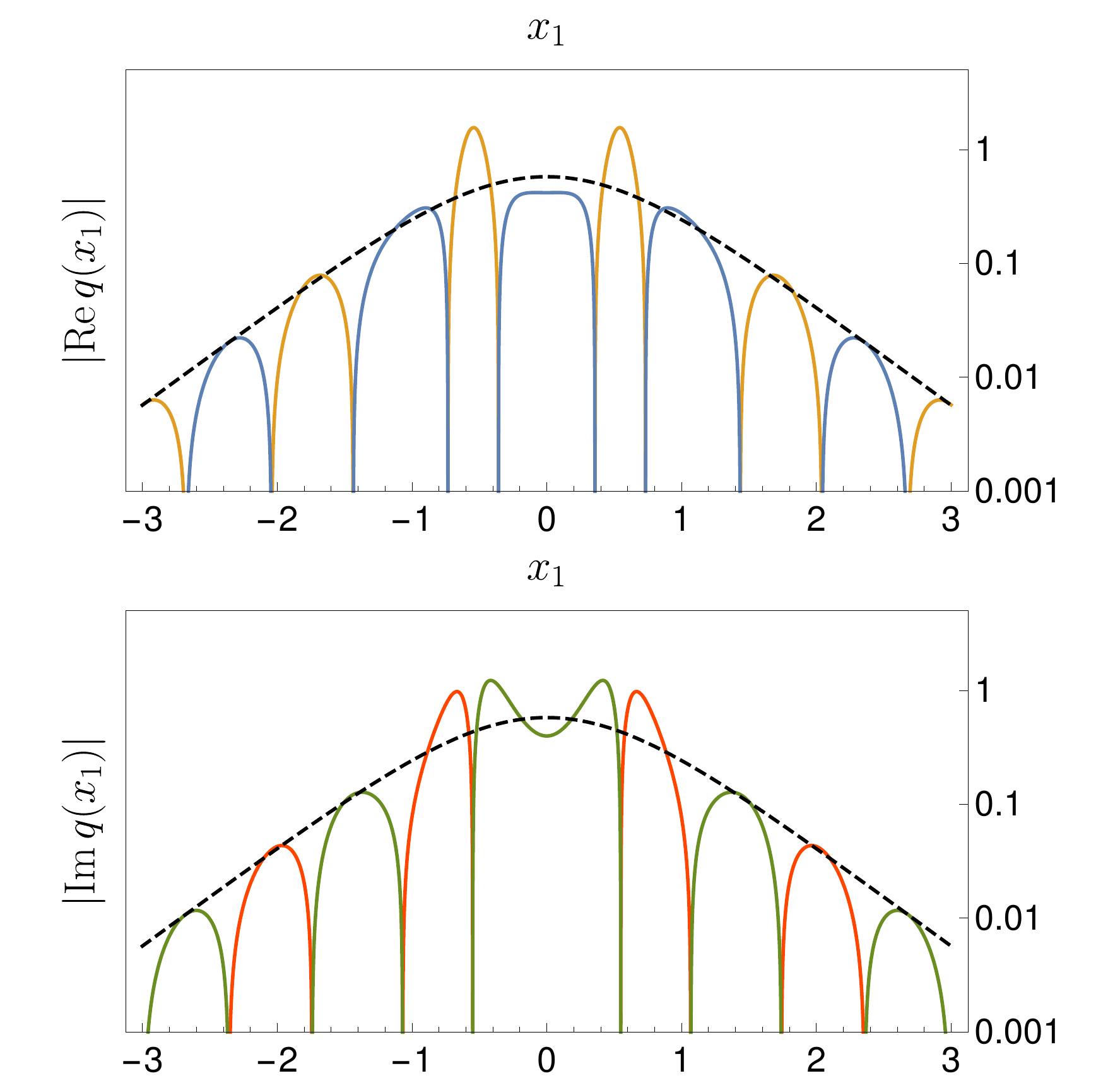}
 \caption{\label{fig:constituent_act_dens}
 Logarithmic plot of $\pm\, \re\, q$ (orange/blue) and $\pm\,
 \im\,q$ (red/green) of the charge density \protect\eqref{eq:constituent_act_dens} with $\Delta x_1=0$ and $\mu=2.5+i$, compared to the density at purely imaginary $\mu'=i$ (i.e., proportional to $1/\cosh^2(x_1)$, dashed grey, multiplied by $|\mu|^2$). The strong dips mark zeros (of just real or imaginary part) at which the corresponding signs (and thus colors) change.}
\end{figure}

Fig.~\ref{fig:constituent_act_dens} shows the behavior of this density for a complex $\mu$ as a function of $x_1$, which is oscillatory and decays exponentially with the imaginary part of $\mu$.

\begin{figure}[!t]
\includegraphics[width=\linewidth]{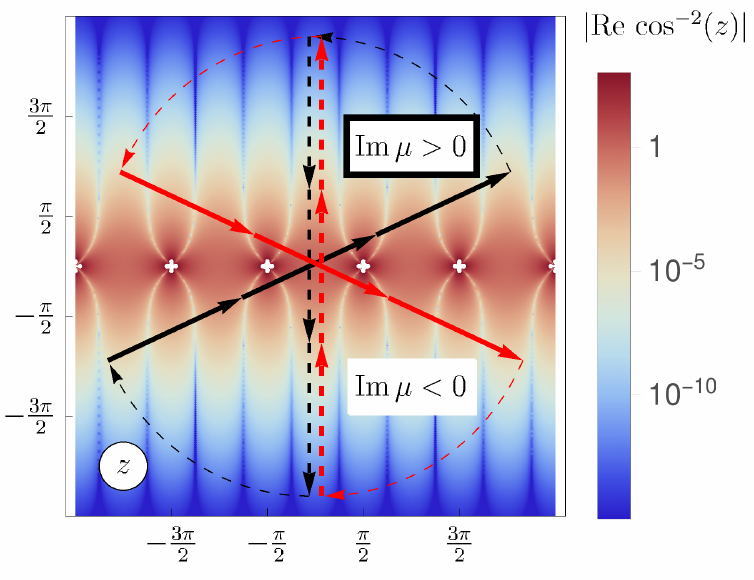}
\caption{Evaluation of the density \eqref{eq:constituent_act_dens} as a contour integral of $\cos(z)^{-2}$ along the tilted  real line $z(x_1)=\mu x_1$, for $\mu = 2.5\pm i$ (black/red) and $\Delta x_1=0$. 
As the infinite arcs do not contribute, the contours can be closed with the imaginary axis, however, with different orientations for $\im\mu \lessgtr 0$, causing the jump in the total charge \protect\eqref{eq:Q_const_o3}. The latter has a singularity at $\im \mu =0$, when the integrand is purely oscillatory and the contour traverses the poles at $z = (2n+1)\pi$.}
\label{fig:contour_O3}
\end{figure}

\begin{figure}[t]
\includegraphics[scale=1.2]{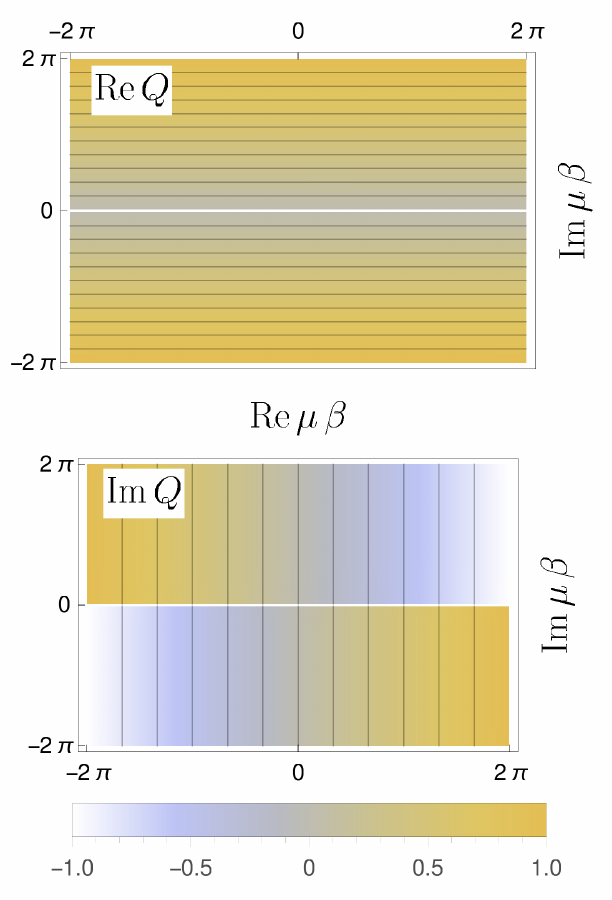}
\caption{Real (top) and imaginary part (bottom) of the total charge $Q$ of a constituent solution as a function of $\mu \beta$ (Eqs. \eqref{eq:Q_const_sep},\eqref{eq:Q_const_o3}): $\im Q$ jumps at $\im \mu= 0$. 
\label{fig:Q_mu_const_o3}}
\end{figure}

In expression \eqref{eq:constituent_act_dens} for $q$, the argument of the cosine describes a contour in the complex plane, in the direction of $\mu$ and parametrized by $x_1$. The corresponding integral can be analyzed by deforming the contour, see Fig.~\ref{fig:contour_O3}.
In any case, for the topological charge and action to be finite, the term $1/\cos^2$ needs a decaying part along this direction, i.e., $\mu$ must possess an imaginary part. 
Further note, that -- as an unavoidable consequence of the complexification of the theory -- there exist `prohibited' values of $\Delta x_1$ for any given $\mu$, for which the denominator in \eqref{eq:constituent_act_dens} vanishes (at some $x_1$) and the topological density is singular (i.e., the contour in Fig. \ref{fig:contour_O3} crosses a pole). 
For any other choice, the total topological charge (and the total action $S=4\pi Q$) can be evaluated
\begin{subequations}
 \begin{align}
 \re Q
 &=\frac{\beta}{2\pi}\,|\im \mu|\,,\\
 \im Q
 &=\frac{\beta}{2\pi}\,\re \mu\cdot\text{sign}(\im \mu)
 \qquad
 (\im \mu\neq 0)\,,
 \end{align}
 \label{eq:Q_const_sep}%
\end{subequations}
as visualized in Fig.~\ref{fig:Q_mu_const_o3},
or, in a more condensed way,
\begin{align}
\label{eq:Q_const_o3}
 Q
 =-\frac{i\mu\beta}{2\pi}\,\text{sign}(\im \mu)
 =\frac{\beta}{2\pi}\,\sqrt{-\mu^2}\quad(\im \mu\neq 0)\,.
\end{align}
where we have made use of the branch cut of the square root function at negative arguments. This again agrees with the twisted case of Eq.~\eqref{eq:mu_omega}. As the topological charge of these solutions is fractional, we will refer to these solutions as constituents. In the CP(N-1) framework we will discuss more general solutions including full instantons made out of such a constituent and another one with complementary charge, see Sec.~\ref{sec:example_cpn}. 

Naively extending the notion of covering to complex target spaces, we can give an interpretation of the complex fractional charge: As $x_2\in[0,\beta]$ and $x_1\in(-\infty,\infty)$, the angles $\pht$ and $\tht$ cover the intervals $[0,i\mu\beta]$ (`fractional') and $[0,\pi]$ (`full'), where the direction of the latter depends on $\im\mu$ giving the sign of $Q$.

For a purely real $\mu$, the topological and action density of these constituent solutions do not decay in space and for $\Delta x_1 \in \mathbb{R}$ even possess infinitely many poles, such that the total topological charge and action integrals are both non-convergent. One may `regularize' this singularity by extending the chemical potential into the complex plane, where, as we have shown, topological charge and action of such  solutions become finite indeed, and then returning to real chemical potentials (such a strategy -- in a complex coupling -- is followed in attempts to define non-alternating series through lateral Borel resummations, which generically produce imaginary parts and jumps, too).
The resulting topological charge and action depend on the way the original real $\mu$ is approached: they receive imaginary parts different on both sides of the real line and as the imaginary part of $\mu$ tends to zero, charge and action are just these imaginary jumps. In the integral of the corresponding densities \eqref{eq:constituent_act_dens} one may understand this as a reversed orientation of the integration contour when crossing the poles, see Fig.\ \ref{fig:contour_O3}.

Other constituent solutions can easily be shown to contain jumps in $Q$ at shifted $\im \mu$: consider twisted solutions with additional Fourier components periodic in $x_2$\footnote{see also the discussion of Fourier components below Eq.~\eqref{eq:wv_four}}: $\ut'=\ut\cdot \exp(2\pi r z/\beta)$ and $\vt^{*\,\prime}=\vt^*\cdot \exp(2\pi r z^*/\beta)$ with $r\in\mathbb{Z}$. This amounts to shifting the chemical potential, $(\ut,\vt^*)'=(\ut,\vt^*)_{\mu\to\mu+2\pi i r/\beta}$, and thus all properties discussed so far apply upon shifting the imaginary part of $\mu\beta$ by a multiple of $2\pi$.

Finally, note that in our solution $\pht=i\mu x_2+\text{const.}$ and $\tht(x_1)$, such that the untwisted angles \eqref{eq:def_twisted2} -- and hence all field representations -- are completely $x_2$-independent. Actually, constant $\ph$'s drop from the Lagrangian which reduces to $2\mathcal{L}_\mu=(\partial_{x_1}\th)^2-\mu^2\sin^2\th$. This is the sine-Gordon quantum mechanics analyzed in \cite{Dunne:2012ae,Fujimori:2016ljw}, but with complex prefactor in the potential term, and $\th(x_1)$ is a (complex) kink. 

\section{CP(N-1) models}

For the family of CP(N-1) models with chemical potential we can repeat the steps done in the O(3) model: incorporating the chemical potential into twisted boundary conditions, complexifying the twisted fields and looking for BPS-like complex solutions. 

\subsection{Preparation: BPS solutions, symmetries, chemical potential and twisted fields}
\label{sec:cpn_prep}

The CP(N-1) field $\n$ is an $N$-dimensional complex vector with $\n^\dagger n=1$. Contact between the lowest nontrivial model CP(1) and the O(3) model, to which it is equivalent, can be made by the transformation $\et_a=\n^\dagger\sigma_a \n$ with $\sigma_{1,2,3}$ the Pauli matrices.

The Lagrangian (at $\mu=0$) can be written down most conveniently by virtue of an auxiliary gauge field:
\begin{align}
 \mathcal{L}_0=(D_\nu \n)^\dagger D_\nu\n\,,\quad
 D_\nu=\partial_\nu+iA_\nu\,,\quad 
 \n^\dagger \n=1\,.
 \label{eq:Lag_cpn}
\end{align}
Since $A_\nu$ enters the Lagrangian quadratically, its equation of motion 
$A_\nu = i\n^\dagger \partial_\nu \n$ can be used at any point in the calculation, e.g., to arrive at an action quartic in $\n$\footnote{Starting with the quartic CP(1) action and using  $\et_a=\n^\dagger\sigma_a \n$ one obtains half the O(3) action \eqref{eq:lagrange_orig1}, which explains the factor of $2$ between the bound $2\pi q$ in Eq.~\eqref{eq:cpn_bound} below and the O(3) bound $4\pi q$ in Eq.~\eqref{eq:o3_bound}.}, but $A_\nu$ can also be treated as an additional field to be path integrated over. This gauge field represents a U(1) gauge invariance $\n\to e^{i\Lambda(x)}\n$ (with the same phase $\Lambda$ for all $\n$-components) under which $A_\nu\to A_\nu-\partial_\nu\Lambda$. One of the $2N-1$ real degrees of freedom in the normalized field can thus be gauged. 

Searching for classical solutions we can again make use of the BPS formalism. Defining complex gauge fields and covariant derivatives,
\begin{align}
\label{eq:compl_der}
 \pbar{A}=(A_1\mp iA_2)/2\,, \quad
 \pbar{D}=(D_1\mp iD_2)/2\,,
\end{align}
the Lagrangian can be written as
\begin{align}
\label{eq:bogo_cpn}
 \la_0 &= 2\left( \para Dn\para^2 + \para \bar{D} n\para^2 \right)\\
 &= 4 \,\para \pbar{D} n \para^2 \mp 2\pi q\,,
\label{eq:bogo_cpn2}
\end{align}
with topological density 
\begin{align}
 q &= \frac{1}{\pi} \left( \para Dn\para^2 - \para \bar{D} n\para^2 \right)\,,
 \label{eq:q_cpn_first}
\end{align}
that is proportional to the curl of the gauge field and thus a total derivative (see Eq.~\eqref{eq:q_cpn} below).
The action thus becomes minimal if
\begin{align}
\label{eq:BPS_cpn}
 \pbar{D}\n_a=0 \quad \forall a\,,
\end{align}
or, for unconstrained fields $\n=\vv/|\vv|$, if
\begin{align}
 \partial^{(*)}\vv_a=0 \quad \forall a\,.
\label{eq_partial_cpn}
\end{align} 
The latter are again solved by (anti)holomorphic functions.

\bigskip

The global U(N) symmetry $\n\to V\n$ with $V\in U(N)$ can be used to define $N-1$ independent conserved charges and chemical potentials. As in the O(3) model, the chemical potentials enter as hermitian terms next to the time derivative,
\begin{align}
 \mathcal{L}_\mu
 =\mathcal{L}_0 \text{ w/ } 
 \begin{array}{l}
 \:\partial_2 \n_a\to (\partial_2-\mu_a)\n_a\\\
 \partial_2 \n^*_a\to (\partial_2+\mu_a)\n^*_a
 \end{array}
 \quad a=1,\ldots,N
\label{eq:lagrange_intro_mu_cpn}
\end{align}
A chemical potential that is the same for all components amounts to a rotation in the U(1) gauge symmetry and therefore has no effect.

We collect the chemical potentials into a diagonal matrix,
\begin{align}
 \M=\text{diag}(\mu_1,\ldots,\mu_N)\,,
\end{align}
and again define twisted fields, 
\begin{subequations}
\begin{align}
 \nt
 &:=e^{-\M x_2}\, \n\,,\\
 \nt^\dagger
 &:=n^\dagger\, e^{\M x_2}
 \qquad(\text{imag.}\,\mu)\,,
 \end{align}
\label{eq:trafo_cpn}%
\end{subequations}
that obey twisted boundary conditions,
\begin{subequations}
 \begin{align}
 \nt(x_2+\beta)&=e^{-\M \beta}\nt(x_2)\,,\\
 \nt^\dagger(x_2+\beta)&=\nt^\dagger(x_2)\,e^{\M \beta}\qquad(\text{imag.}\,\mu)\,.
 \end{align}
\label{eq:bcs_cpn}%
\end{subequations}
The same relation between Lagrangians with and without $\mu$ applies as before in Eq.~\eqref{eq:relating_lagrangians},
\begin{align}
 \mathcal{L}_\mu(\n)
 =\mathcal{L}_{\mu=0}(\nt)\,.
\end{align}

However, the problem encountered in the O(3) model also occurs here: for generic $\mu$ the definitions in Eq.~\eqref{eq:trafo_cpn} and consequently the boundary conditions Eq.~\eqref{eq:bcs_cpn} are \textit{not compatible with complex conjugation}, unless $\mu$ is purely imaginary. For the latter, solutions can be obtained by means very similar to those in O(3).

\subsection{Complexification and complex solutions}
\label{sec:comp_comp}

Looking for saddles at nonzero chemical potential, we first have to \textit{complexify the fields}. As in the O(3) model we do this by treating the field $\nt$ and its conjugate as independent, renaming $\nt^\dagger\to\mt^\dagger$, that are subject to the constraint 
\begin{align}
 \mt^\dagger \nt=1\,,
 \label{eq:constraint_cpn}
\end{align}
and to twisted boundary conditions
\begin{subequations}
 \begin{align}
 \nt(x_2+\beta)&=e^{-\M \beta}\nt(x_2)\,,
 \label{eq:bcs_cpn_new_1}\\
 \mt^\dagger(x_2+\beta)&=\mt^\dagger(x_2)\,e^{\M \beta}\,.
 \end{align}
\label{eq:bcs_cpn_new}%
\end{subequations}
For the new Lagrangian generalizing \eqref{eq:Lag_cpn} one may allow for an independent auxiliary gauge field in the derivative of $\mt^\dagger$, but by its equation of motion it equals the complex conjugate of $A_\nu$. Eventually we find
\begin{align}
\label{eq:Lag_mn}
 \mathcal{L}_0(\mt,\nt)
  &=\big[\DD_\nu\mt\big]^\dagger
 D_\nu\nt\,,\qquad
 \DD_\nu=\partial_\nu+i A_\nu^*\\
 &=-\mt^\dagger D_\nu^2\nt\,,
\end{align}
where the auxiliary gauge field
\begin{align}
\label{eq:A_compl}
 A_\nu =i\mt^\dagger\partial_\nu \nt\,,
\end{align}
is now complex as well (and periodic). It is related to a complexified gauge symmetry $GL(1,\mathbb{C})$
\begin{align}
 \nt\to e^{i\Xi(x)}\nt\,,\quad
 \mt^\dagger\to \mt^\dagger e^{-i\Xi(x)}\,,\quad
 A_\nu\to A_\nu-\partial_\nu \Xi
 \label{eq:cpn_comp_local}
\end{align}
with $\Xi$ a complex number (field). This and the complex constraint \eqref{eq:constraint_cpn} remove four of the $4N$ real degrees of freedom, which are therefore twice the $2N-2$ real degrees of freedom of the real case. 

In the same way the global symmetry is extended to $GL(N,\mathbb{C})$,
\begin{align}
 \nt\to\mathcal{V}\nt\,,\quad
 \mt^\dagger \to\mt^\dagger\mathcal{V}^{-1}\,,\quad
 \mathcal{V}\in GL(N,\mathbb{C})\,.
 \label{eq:cpn_comp_global}
\end{align}

The complex nature of the gauge field also requires a modified definition of the complexified covariant derivative (completing \eqref{eq:compl_der}) : 
\begin{subequations}
\label{eq:mod_compl_der}
\begin{align}
 \pbar{D} 
 &= (D_1 \mp i D_2 )/2 
 = \pbar{\partial} + i\pbar{A}\,,\\
 \DD
 &= (\DD_1-i\DD_2)/2
 =\partial+i \bar{A}^*\,,\\
 \bar{\DD}
 &= (\DD_1+i\DD_2)/2
 =\bar{\partial}+i A^*\,,
\end{align}
\end{subequations}
where $\bar{\partial} = \partial^*$ (see \eqref{eq:complex_coord}), but in the gauge field we have to distinguish between combinations of $A_1$ and $iA_2$ denoted by a bar and complex conjugation denoted by an asterisk. To be fully clear we list all four quantities
\begin{align}
 A
 &=(A_1-iA_2)/2\,,\qquad\:\,
 \bar{A} 
 =(A_1 +iA_2)/2\,,\\
 A^*
 &=(A_1^*+i A_2^*)/2\,,\qquad
 \bar{A}^*
 =(A_1^* -iA_2^*)/2\,.
\end{align}
Only for real $A_{1,2}$ one has $A^*=\bar{A}$ and $\bar{A}^*=A$
and $\DD=D$ (and $\DD_\nu=D_\nu$). The modified covariant derivatives satisfy $D^\dagger = - \bar{\DD}$ and $\bar{D}^\dagger = -\DD$. With these definitions, the Lagrangian is conveniently written as
\begin{align}
 \la_0 
 &= 2\left(
 [\DD \mt]^\dagger D\nt
 +[\bar{\DD} \mt]^\dagger \bar{D}\nt
 \right)\\
 &=-2\,\mt^\dagger\left(\bar{D}D+D\bar{D} \right)\nt\,,
\end{align}
and rewritten as
\begin{align}
 \la_0 
 &= \begin{cases}
 4[\bar{\DD} \mt]^\dagger \bar{D}\nt + 2\pi q \\
 4[\DD \mt]^\dagger D\nt - 2\pi q\,, \\
 \end{cases} 
 \label{eq:cpn_bound}%
 \end{align}
with `topological density'
\begin{align}
 q &= \frac{1}{\pi} \left([\DD
 \mt]^\dagger D\nt - [\bar{\DD} \mt]^\dagger \bar{D}\nt \right)\,.
 \label{eq:q_again}
\end{align}
These formulas are the analogues of the real case expressions \eqref{eq:bogo_cpn}-\eqref{eq:q_cpn_first} (which they become for $\mt \rightarrow \nt$ and $\DD\to D$). The topological density\footnote{The operator $[D,\bar{D}]$ in \eqref{eq:q_again} is originally sandwiched between $\mt^\dagger$ and $\nt$, but since it is a scalar, it is just a product with $\mt^\dagger\nt=1$. The analogous commutator $[\DD,\bar{\DD}]$ relates $q^*$ to the curl of $A^*$. \label{foot:sandwich}}
\begin{align}
\label{eq:q_cpn}
 q = \frac{1}{\pi}\,[D,\bar{D}] 
 =  \frac{i}{2\pi}\,[D_1,D_2] 
 = \frac{1}{2\pi}\,\epsilon_{\mu\nu} \partial_\nu A_\mu\,,
\end{align}
is now generically complex -- precluding the application of the conventional BPS-argument -- but still a total derivative\footnote{
The topological density is also $GL(N,\C)$ invariant and periodic.}. As such it does not contribute to the equations of motion for $\mt^\dagger$ and $\nt$, which are derived in the usual way incorporating the constraint \eqref{eq:constraint_cpn} by a Lagrange multiplier, resulting in
\begin{subequations}
\begin{align}
 D\bar{D}\nt \;-\;\, \left[\mt^\dagger(D\bar{D}\nt) \right]\nt &=0\, ,\\
 (\DD \bar{\DD}\mt)^\dagger - \Big[(\DD \bar{\DD} \mt)^ \dagger \nt\Big]  \mt^\dagger &= 0\,.
 \end{align}
\label{eq:eom_cpn_mod}%
\end{subequations}
Note that in these equations one may commute $D$ with $\bar{D}$ and $\DD$ with $\bar{\DD}$ (in both terms simultaneously\footnote{since the commutators are scalars and drop out of the particular projectors in the equations of motion}). These equations are again easily solved by vanishing complex covariant derivatives, $\pbar{D}\nt=0$ and $\pbar{\DD}\mt=0$, for all components. 

Furthermore, the variational principle implies that for stationary points\footnote{For the real model, this follows from a finite action requirement for \emph{all} configurations. This no longer holds in the complexified theory.}, \emph{all} complex covariant derivatives must vanish at the spatial boundaries, i.e. 
\begin{align}
 |x_1|\to\infty:\quad
 \para \pbar{D} \nt \para \rightarrow 0  \,,\quad
 \para \pbar{\DD} \mt \para \rightarrow 0,
\end{align}
forcing the fields to assume the asymptotic forms (modulo a gauge transformation)
\begin{align}
 \label{eq:mn_asympt}
 x_1\to\pm\infty:\quad
 \nt 
 &\rightarrow e^{\,i\,\Xi_\pm(x_2)}{n}_\pm\,,\notag\\
 \mt^\dagger 
 &\rightarrow e^{-i\,\Xi_\pm(x_2)}{m}_\pm^\dagger\,, 
\end{align}
i.e. an $x_2$-dependent element of the $GL(1,\C)$ gauge symmetry (times a constant vector), individually for each boundary (the derivation is analogous to that for the real CP(N-1) model, see  e.g.\ \cite{Rajaraman:1982is}). This establishes an asymptotic relation between the a priori independent fields $m^\dagger$ and $n$, and provides the basis for our topological discussions below (Sec. \ref{sec:topol}).

We can again construct unconstrained fields as
\begin{align}
\label{eq:mn_unconst}
 \nt
 =\frac{\vvt}{\sqrt{\wt^\dagger \vvt}}\,,\qquad
 \mt^\dagger
 =\frac{\wt^\dagger}{\sqrt{\wt^\dagger \vt}}\,,
\end{align}
such that $\mt^\dagger\nt=1$ indeed. The action density in these unconstrained variables is
\begin{align}
 \mathcal{L}_0
 =\frac{\partial_\nu\wt^\dagger}{\sqrt{\wt^\dagger \vt}}
 \,P\,
 \frac{\partial_\nu\vt}{\sqrt{\wt^\dagger \vt}}\,,\qquad
 P=\Big(1_N-\frac{\vt\wt^\dagger}{\wt^\dagger \vt}\Big),
\end{align}
where $P$ is a projector perpendicular to $\vt$ and $\wt^\dagger$: $P\vt=0=\wt P$. 

Again, the equations of motion \eqref{eq:eom_cpn_mod} are fulfilled, if the BPS-like conditions $\partial^{(*)}\vt=0$ and $\partial^{(*)}\wt=0$ are obeyed (as in \eqref{eq_partial_cpn}, but separately for $\vt$ and $\wt$). The four sets of solutions then amount to
\begin{subequations}\begin{align}
 \vt=\vt(z)\: \quad 
 \wt&=\wt(z)\: \qquad \: (\mathcal{L}_0=+2\pi q)
 \,,
 \label{eq_the_solutions_cpn_1}\\
 \vt=\vt(z^*) \quad 
 \wt&=\wt(z^*) \qquad (\mathcal{L}_0=-2\pi q)
 \,,
 \end{align}
\label{eq_the_solutions_cpn_first}%
\end{subequations} 
or 
\begin{subequations}
 \begin{align}
 \vt=\vt(z)\: \quad 
 \wt&=\wt(z^*)\: \qquad (\mathcal{L}_0=0=q)
 \,,\\
 \vt=\vt(z^*) \quad 
 \wt&=\wt(z) \qquad \:\:\,(\mathcal{L}_0=0=q)
 \,,
 \end{align}
\label{eq_the_solutions_cpn_second}%
\end{subequations}
as in the O(3) model, Eqs.~\eqref{eq_the_solutions_o3_first} and \eqref{eq_the_solutions_o3_second}. For these BPS-like 
solutions, gauge fields and charge density only depend on derivatives of $\log (\wt^\dagger \vt)$:
\begin{align}
 \pbar{A} = \pm \frac{i}{2} \pbar{\partial} \log(\wt^\dagger \vt)\,, \qquad
 q = \pm \frac{1}{4\pi} \Delta \log(\wt^\dagger \vt),
 \label{eq:logs}
\end{align}
analogous to the well-known real case.

\bigskip

A generic (anti)holomorphic solution satisfying the boundary condition \eqref{eq:bcs_cpn_new} is specified by two Fourier series
\begin{subequations}
\begin{align}
\vt(z^{(*)}) = e^{\pm iMz^{(*)}} \sum_{r=r_-}^{r_+} a^{(r)} \, e^{\frac{2\pi}{\beta} r z^{(*)}} \\
\wt(z^{(*)}) = e^{\mp iM^\dagger z^{(*)}} \sum_{s=s_-}^{s_+} b^{(s)}\, e^{\frac{2\pi}{\beta} s z^{(*)}}
\end{align}
\label{eq:wv_four}%
\end{subequations}
with vector valued coefficients $a^{(r)},\, b^{(s)} \;\in \C^N$. Despite the general ansatz, these expansions are \emph{not independent}, as shown in the following. Without loss of generality\footnote{Any solution for $\mu_a$ is equivalent to a solution for $\mu^\prime_a = \mu_a - i\frac{2\pi}{\beta} k_a, \; k_a\in\mathbb{Z}$ with a different periodic part, obtained by shifting the coefficients $a^{(r)}_a \rightarrow a^{(r \pm k_a)}_a $ and $b^{(s)}_a \rightarrow b^{(s \pm k_a)}_a $, for the (anti)holomorphic cases, respectively.} we can restrict $\im \mu_a \in(-2\pi/\beta,0]$. The asymptotic form of \eqref{eq:wv_four} for the (exemplary) holomorphic case at $x_1 \rightarrow \pm \infty$ then reads
\begin{subequations}
\begin{align}
 \vt(z) \; &\rightarrow \;  e^{iMz} a^{(r_\pm)}\;e^{\frac{2\pi}{\beta} r_\pm\, z},\\
 \wt^\dagger(z^*) \; 
 &\rightarrow \; (b^{(s_\pm)})^\dagger \, e^{iMz^*} \;e^{\frac{2\pi}{\beta} s_\pm\, z^*}
 \label{eq:vw_asympt}
\end{align}
\end{subequations}
The term $e^{-\im M x_1}$ projects onto some direction $a_\pm$ and $b_\pm$, respectively, determined by the imaginary parts of the eigenvalues $\mu_a$. The dominating $\mu$'s at $x_1 \rightarrow \pm \infty$ are denoted by $\mu_\pm$. Then,
\begin{align}
 &\wt^\dagger \vt (x_1,x_2) \; \rightarrow \label{eq:wv_asymptot}\\ 
 &\qquad(b_\pm^\dagger a_\pm) \;\cdot\; e^{ \frac{2\pi}{\beta}(r_\pm + s_\pm + i\mu_\pm \beta/\pi) \,x_1}
 \cdot\, e^{i \, \frac{2\pi}{\beta} \,(r_\pm -s_\pm) \, x_2 }\,.\nonumber
\end{align}
Demanding finite fundamental fields $\nt, \,\mt^\dagger$ one needs to impose $\wt^\dagger \vt \neq 0$, according to the parametrization \eqref{eq:mn_unconst}. This establishes a necessary (but not sufficient) condition for the powers in the expansions \eqref{eq:wv_four}: since $\wt^\dagger \vt(x_1,x_2)$ is periodic in $x_2$, the (integer) quantity
\begin{align}
K(x_1) := \frac{-i}{2\pi}\int_0^\beta \partial_2 \log \wt^\dagger \vt \, dx_2 
\end{align}
counts the winding of $\wt^\dagger \vt$ around the origin, for any fixed $x_1$. It jumps if (and only if) $\wt^\dagger \vt$ crosses the origin. Avoiding a divergence of $\nt,\, \mt^\dagger$ in the bulk therefore requires a \emph{constant} $K(x_1) = K \in \mathbb{Z}$.
Comparison with the asymptotics \eqref{eq:wv_asymptot} yields
\begin{align}
K(x_1 \rightarrow \pm \infty) = r_\pm -s_\pm
\end{align}
and thus $r_+ -s_+ = r_--s_-$, or
\begin{align}
r_+-r_- = s_+-s_-.
\end{align}
This condition effectively equates the dimensions of the moduli spaces of the two `partial solutions' $v$ and $w^\dagger$.
Furthermore, the vectors $a^{(r_\pm)}$ and $b^{(s_\pm)}$ must be chosen such that $b_\pm^\dagger a_\pm \neq 0 $ in \eqref{eq:wv_asymptot}. If additionally one requires the fields to belong to the original (not complexified) field manifold in the limit of purely imaginary (or vanishing) $\mu$, all coefficients in \eqref{eq:wv_four} must be chosen identical, $a_a^{(r)}=b_a^{(r)}$. 

\subsection{Topology}
\label{sec:topol}
The topological charge, as defined in Eq. \eqref{eq:q_cpn} in analogy to the real case, is directly related to the difference $\Delta\Xi := \Xi_+-\Xi_-$:
\begin{subequations}
\begin{align}
 2\pi Q 
 &= -\int_{\partial(\mathbb{R}\times \mathbb{S}^\beta)} A_\nu d\sigma_\nu 
 = - \int_0^\beta dx_2 A_2\big|_{x_1\rightarrow -\infty}^{x_1\rightarrow +\infty} \\
 &= \Delta\Xi(\beta) -\Delta\Xi(0),
 \end{align}
\label{eq:Q_surf}%
\end{subequations}
where we have used the periodicity of the gauge field and its asymptotic form defined by Eqs.\ \eqref{eq:mn_asympt} and \eqref{eq:A_compl}.
The real part of $Q$, corresponding to the compact direction of $GL(1,\C)$, retains its interpretation as a winding number from the real model. In particular, \textit{non-twisted} solutions of the complexified model, which must satisfy $\Delta\Xi(\beta) = \Delta\Xi(0) + 2\pi k$, still exhibit an integer charge $Q=k$, associated now with $\pi_1(GL(1,\C)) = \mathbb{Z}$. Any fractional (real) and/or imaginary contribution to $Q$ must therefore be rooted entirely in the twist, i.e., in the chemical potential.
For that we combine the asymptotic behavior \eqref{eq:mn_asympt} with the twisted boundary condition \eqref{eq:bcs_cpn_new} to $ e^{\,i\,\Xi_\pm(\beta)}\cdot {n}_\pm = e^{\,i\,\Xi_\pm(0)}\cdot e^{-M\beta} {n}_\pm$ and similarly for ${m}_\pm$ (as $m_\pm^\dagger\, n_\pm=1$). Seen as a vector equation this implies that ${n}_\pm$ are eigenvectors of $e^{-M\beta}$. The corresponding eigenvalues are of the form $e^{-\mu_\pm \beta}$ where $\mu_\pm\in\{\mu_1,\ldots\mu_N\}$, which eigenvalue is assumed is determined by the asymptotics of $n$ (see also below \eqref{eq:vw_asympt} for BPS-like solutions). It follows that $\Xi_\pm(\beta)-\Xi_\pm(0)=i\mu_\pm\beta$ modulo $2\pi$ and with 
$\Delta \mu := \mu_+ - \mu_- $ the topological charge becomes
\begin{align}
 \!\!\!\!\! Q 
 =k+i \, \frac{\beta}{2\pi}\,\Delta\mu
 = (k -\frac{\beta}{2\pi}\, \im\Delta \mu) 
 + i \,\frac{\beta}{2\pi}\, \re \Delta\mu. 
\end{align}
If $\mu_+ = \mu_-$ the charge is again integer. This is the case in particular if $n_+ = n_-$, in turn \emph{a non-integer charge requires the solution to interpolate between different field components}. The reverse is only true for mutually distinct $\mu_a$.

\begin{figure}[!t]
 \includegraphics[width = \linewidth]{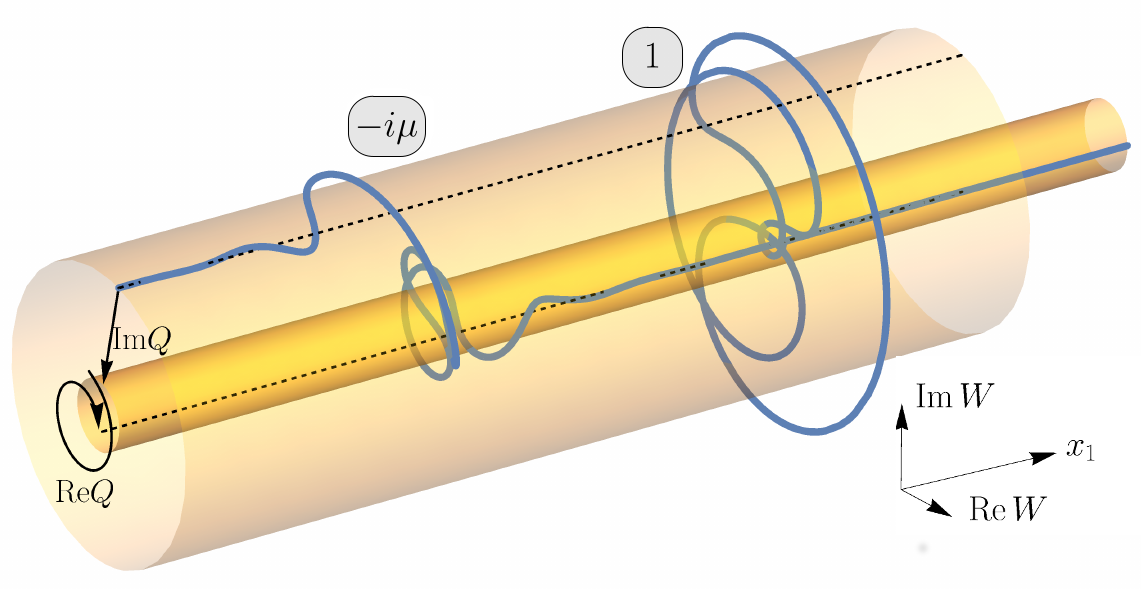}
 \includegraphics[width = \linewidth]{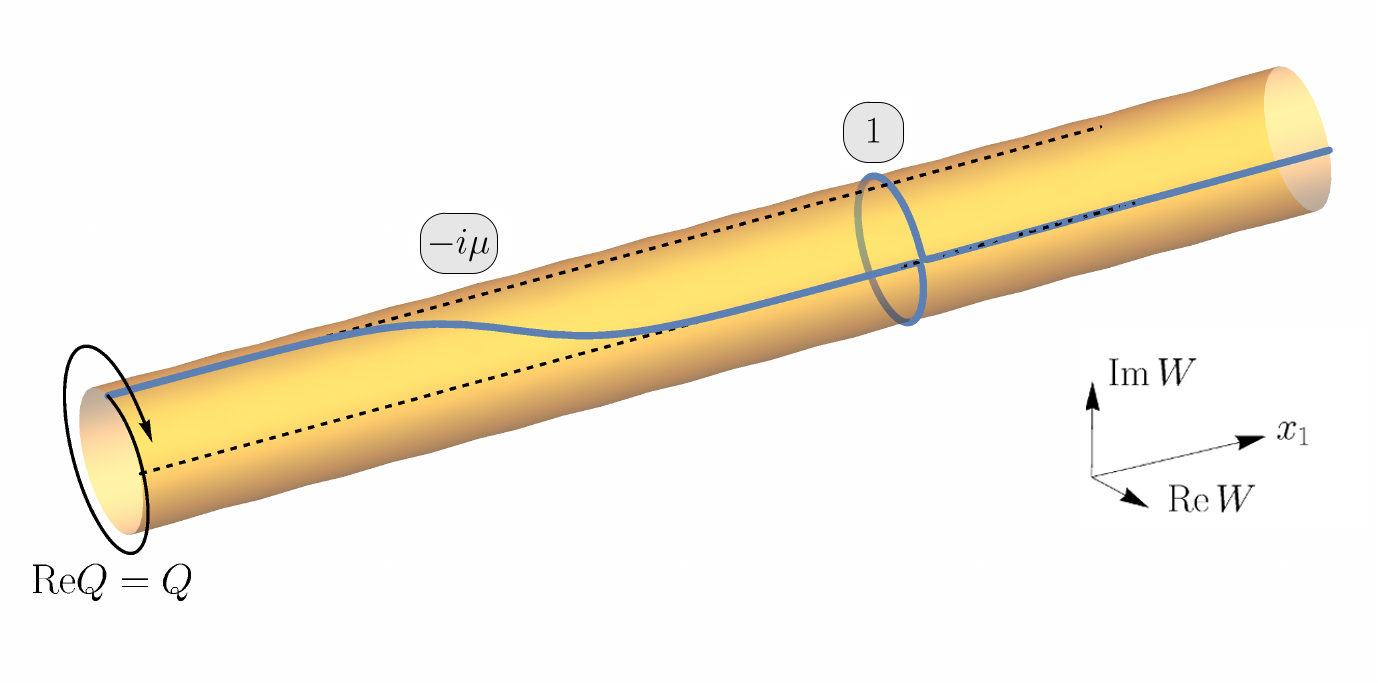}
 \caption{Top: Polyakov line \eqref{eq:Wilson} of the `full instanton' solution from Sec. \ref{sec:example_cpn}, for $\mu\beta = 2\pi( 0.5 + 0.2i)$ (similar to case (b) in Figs.\ \ref{fig:intersection}-\ref{fig:qim}). $\re Q$ and $\im Q$ are given by the difference of the asymptotic values (as $x_1 \rightarrow \pm \infty$) of its angular (compact) and radial (non-compact) component, respectively. Two constituents can be recognized, where $W(x_1)$ changes rapidly. Only the left one contributes to $\im Q$, while the right one has charge $Q=1$ ($W$ essentially winds around the cylinder once). Bottom: Polyakov line for purely imaginary $\mu\beta = 2\pi \cdot 0.2i$. The constituents are still visible, but the radial component of $W$ is constant.}
 \label{fig:Wilson}
 \end{figure}

The topological interpretation of a complex charge can be illustrated nicely by virtue of the Polyakov line
\begin{align}
\label{eq:Wilson}
 W(x_1) = \exp{\left(i\int_0^\beta \!dx_2\, A_2(x_1,x_2)\right)}
\: \in GL(1,\C)\,.
\end{align}
Then $Q = i\left[ \log W\right]_{x_1\rightarrow -\infty}^{x_1\rightarrow+\infty}$ has the interpretation of an accumulated change of $W$ as $x_1$ goes from $-\infty$ to $+\infty$. Fig.\ \ref{fig:Wilson} shows both the fractional covering of the compact direction of the gauge group -- reflected in a fractional (real) contribution to $Q$ -- as well as the extension of the trajectory to a noncompact dimension, resulting in an imaginary contribution to $Q$ when $\re \Delta\mu \neq 0$.

\subsection{Example: full instanton}
\label{sec:example_cpn}

As an example we will now discuss a full instanton at generic $\mu$, which -- roughly speaking -- consists of two constituents of the form discussed in Sec.~\ref{sec:example_o3} rendering it time-dependent. We follow and slightly extend the approach in \cite{Brendel:2009mp} and restrict ourselves to CP(1), similar discussions apply to higher CP(N-1)'s. With the choice of chemical potentials $\mu_2=-\mu_1=-\mu/2$, where $\mu$ is the chemical potential discussed in the O(3) model (and a constant off-set in $\mu_2$ and $\mu_1$ has no effect as discussed below \eqref{eq:lagrange_intro_mu_cpn}), particular solutions read
\begin{subequations}
 \begin{align}
  \vvt(z)
  &=\left(\begin{array}{l}
  e^{-i\frac{\mu}{2}z}\,\big[1+\lambda_2\, e^{2\pi z/\beta}\big]\\
  e^{i\frac{\mu}{2}z}\,\lambda_1
  \end{array}\right)\,,\\
  \wt^\dagger(z^*)
  &=\Big(
  e^{-i\frac{\mu}{2}z^*}\,\big[1+\lambda_2^*\, e^{2\pi z^*/\beta}\big]\,,\:
  e^{i\frac{\mu}{2}z^*}\,\lambda_1^*\Big)\,.
 \end{align}
 \label{eq:vt_wt}%
\end{subequations}
They are of the BPS-like form of Eq.~\eqref{eq_the_solutions_cpn_1} and do obey the required boundary conditions \eqref{eq:bcs_cpn_new}. Moreover, the Fourier coefficients $a$ and $b$ in \eqref{eq:wv_four} have been chosen identical. The absence of the $\exp(2\pi z^{(*)}/\beta)$-term in the second component amounts to a choice of gauge. The parameters $\lambda_{1,2}$ are moduli of the solution, among which $\lambda_2>0$ can be achieved by shifting $x_2$ accordingly.

Using formula \eqref{eq:logs}, the action/topological density is the Laplacian of the logarithm of\footnote{up to a common prefactor $e^{i\mu x_1}$, which corresponds to a gauge transformation and drops out after taking the Laplacian}
\begin{align}
 \wt^\dagger\vvt
 &=e^{\,0\cdot x_1}
 +2\lambda_2\cos(2\pi x_2/\beta)e^{\,2\pi/\beta\cdot x_1}
 +\lambda_2^2\,e^{\,4\pi/\beta\cdot x_1}\notag\\
 &+|\lambda_1|^2 e^{2\, i\,\re \mu\, x_1}
 e^{-2 \,\im \mu\cdot x_1}
 \label{eq:wv}
\end{align}
where we have emphasized four $x_1$-exponents. The interpretation of such a solution relies on the fact that if one of the exponents dominates, the logarithm becomes a linear function in $x_1$ and $q$ vanishes. Therefore, the topological charge is concentrated near points where two of the four terms above intersect (see below). From the same formula it follows that the charge of these lumps is given by the difference of the slopes of intersecting graphs. Note that this statement concerns the full complex charge/action: once the real parts of the exponents have determined the dominating terms/graphs, the full (complex) prefactor of $x_1$ from the dominating exponent enters $\log(\wt^\dagger\vvt)$ and thus the charge. Likewise, the total topological charge is given as the difference of the dominating exponents for $x_1\to\pm\infty$ (see also Eq.~\eqref{eq:Q_surf}).

In this example, we focus on the effect of a varying chemical potential $\mu$ on this solution.
The first three terms in \eqref{eq:wv} are thus considered fixed in this context. Among them, the first, constant term dominates at $x_1\to -\infty$, while the third term dominates at $x_1\to +\infty$, which would determine a total charge $Q=(4\pi-0)/4\pi=1$. The second term gives rise to an $x_2$-dependence in the charge density, localized at the single common intersection point $x_1=-\beta/2\pi\cdot\log \lambda_2$.

\begin{figure}[!t]
 \includegraphics[scale=1.2]{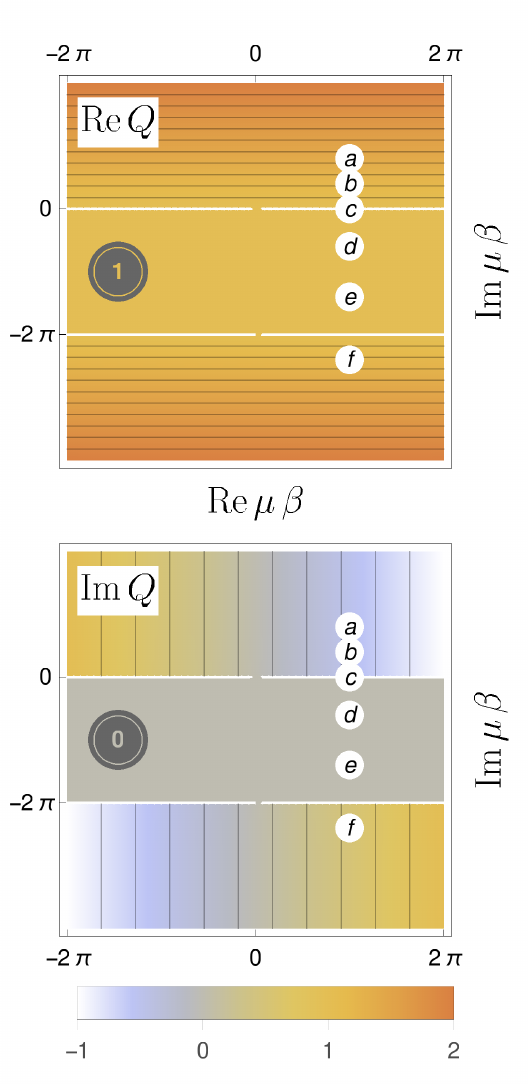}
 \caption{Real (top) and imaginary part (bottom) of the total charge $Q$ of the `full instanton' solution (Eqs.\ \eqref{eq:Q_1}, \eqref{eq:Q_2} or \eqref{eq:Q_full_cpn}) as a function of $\mu\beta$. In the region $\im \mu\beta \in (-2\pi,0)$, the charge is constant, $Q=1$, while at its boundaries, $\im Q$ is discontinuous. 
 Note the similarities to Fig.~\ref{fig:Q_mu_const_o3}.
 The specific values (a) to (f), $\im \mu \, \beta = 2\pi \times\{0.4,0.2,0,-0.3,-0.7,-1.2 \}$ and $\re\mu \beta= 0.8\,\pi$, correspond to the cases depicted locally in Figs.\ \ref{fig:intersection}-\ref{fig:qim} below.}
\label{fig:Q_mu}
\end{figure}

The slope  $-2\,\im \mu$ of the fourth contribution can now compete with 0 or $4\pi/\beta$, and thus affect both the asymptotic behaviour  and number/location(s) of intersection points: for intermediate $\im \mu\in(-2\pi/\beta,0)$, the total charge remains $Q=1$, see also  Fig.~\ref{fig:Q_mu}. For purely imaginary $\mu$ (with $\omega\in[0,1]$) this is the case discussed in \cite{Brendel:2009mp}. Depending on the prefactors, there are one or two intersection points, i.e., the topological density comes in one (time-dependent) or two lumps (`constituents'), see the cases (d) vs.\ (e) below.

\begin{figure*}[!t]
 \includegraphics[width=\textwidth]{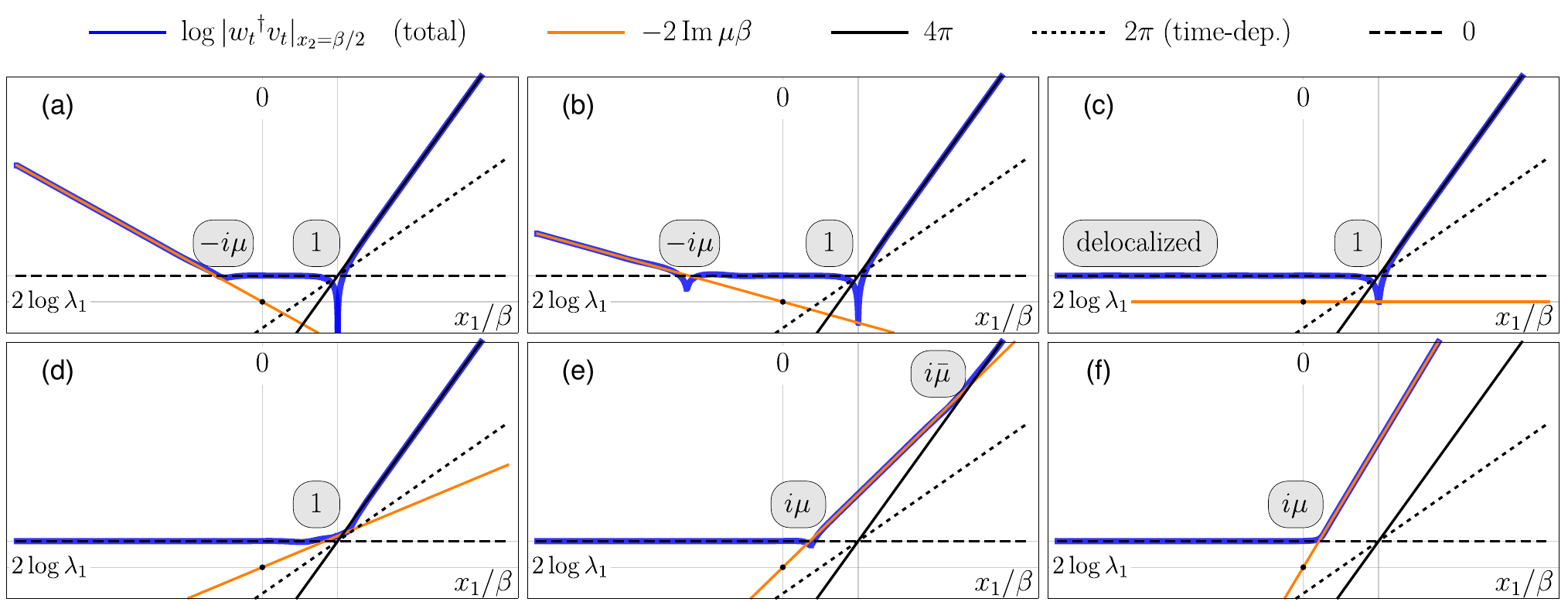}
\caption{Intersection picture for the topological charge $q$ of the `full instanton' solution \eqref{eq:vt_wt}: $\log|\wt^\dagger \vt| $ given in \eqref{eq:wv} (plotted here for fixed $x_2 = \beta/2$) has piecewise dominating exponents and consequently $q$ is localized in lumps at the intersection points of their (linear) graphs. Their number and location(s) varies with $\im\mu$, from (a) to (f): $\im \mu \, \beta = 2\pi \times\{0.4,0.2,0,-0.3,-0.7,-1.2 \}$ and $\re\mu \beta= 0.8 \pi$ (as in Fig.\ \ref{fig:Q_mu}) with common parameters $\lambda_1 = 10^{-1}, \lambda_2 = 10^{-4}$. Charges of these lumps are either integer $Q=1$ or fractional $Q= \pm i \pbar{\mu}$ times $\beta/2\pi$ (the latter factor neglected in the figure labels). At  $\im\mu \beta = 0$ (transition (b) $\leftrightarrow$ (d) at (c)) and $\im \mu \beta = -2\pi$ (transition (e) $\leftrightarrow$ (f)), intersection points and the corresponding topological lumps (dis-)appear at $x_1/\beta =\mp \infty$. The strong dips associated with the unit 
charged 
instanton in (a)-(c) are due to the time-dependent contribution in \eqref{eq:wv} (which is maximally negative for $x_2= \beta/2$).}
\label{fig:intersection}
\end{figure*}

For $\im \mu>0$, the behavior at $x_1\to-\infty$ is changed and the total charge becomes $Q=(4\pi-2i\beta\mu)/4\pi$, likewise for $\im \mu<-2\pi/\beta$ the behavior at $x_1\to+\infty$ is changed and $Q=(2i\beta\mu-0)/4\pi$.
This qualitative difference is also reflected in the asymptotic form of the fields themselves and the corresponding eigenvalues $\mu_\pm$ of $M$ (cf.\ Sec.\ \ref{sec:topol}):
\begin{align}
\!\!\!\!\!\!\begin{cases}
 n_+ =  \ve{1}{0},\,
 n_- = \ve{0}{1},
 &\!\!\!\! \mu_\pm = \pm \frac{\mu}{2}\,   
 \text{ for } \im\mu\beta <-2\pi, \\[4pt]
 n_\pm = \ve{1}{0},
 &\!\!\!\! \mu_\pm = - \frac{\mu}{2}\, 
 \text{ else,}  \\[4pt]
 n_+ = \ve{0}{1},\,
 n_- = \ve{1}{0},
 &\!\!\!\!  \mu_\pm = \mp \frac{\mu}{2}\,  
 \text{ for }\im\mu\beta > 0,
\end{cases}
\end{align}
(and similar for $\mt^\dagger$).
Only in the first and last case, where $n_+ \neq n_-$, can there exist a non-integer contribution to the total charge, which can be summarized as
\begin{align}
 Q 
 &=i\,\frac{\beta}{2\pi}\big(\mu\,\Theta(-\im \mu)+
 \bar{\mu}\,\Theta(-\im \bar{\mu})\big)
 \label{eq:Q_1}\\
 &=\begin{cases}
   {\displaystyle i\,\frac{\mu\beta}{2\pi}} &\text{ for } \im\mu\beta <-2\pi\,,\\
   {\displaystyle 1} & \text{ else,}\\
   {\displaystyle i\,\frac{\bar{\mu}\beta}{2\pi} \;\left(\,=1-i\,\frac{\mu\beta}{2\pi}\right)}&\text{ for } \im\mu\beta >0\,, 
  \end{cases}
 \label{eq:Q_2}
\end{align}
with $\Theta$ the Heaviside function and
\begin{align}
 \bar{\mu}:=-i\,\frac{2\pi}{\beta} - \mu
\label{eq:new_mu} 
\end{align}
playing the role of a complementary chemical potential (for this solution)\footnote{For purely imaginary $\mu$, Eq.~\eqref{eq:mu_omega}, this definition is compatible with the complementary twist $\bar{\omega} = 1-\omega$ defined in \cite{Bruckmann:2007zh}.}.
In a condensed way,
\begin{align}
 Q
 &=\frac{1}{2}+\frac{\beta}{4\pi}\big(\sqrt{-\mu^2}+\sqrt{-\bar{\mu}^2}\big)\,.
 \label{eq:Q_full_cpn}
\end{align}
Here and in \eqref{eq:Q_1} the total charge of this solution can be recognized as the sum of two constituents from the lowest $O(3)=CP(1)$ model, cf.~Sec.\ \ref{sec:example_o3} and in particular Eq.~\eqref{eq:Q_const_o3} with $\mu$ and $\bar{\mu}$, respectively. Note, however, that the Heaviside function in \eqref{eq:Q_1} differs from the sign function in \eqref{eq:Q_const_o3} and that the square roots in \eqref{eq:Q_full_cpn} come with half the prefactor and an additional `interaction' term 1/2. As a result, $Q$ in  \eqref{eq:Q_2} is either the charge of one constituent or unity.

This constituent picture will essentially be validated now looking at the behavior of the topological density. In particular, the jumps in the total charge coincide with constituents (dis-)appearing at spatial infinity.

\renewcommand{\topfraction}{0.85}
\begin{figure*}[!t]
 \includegraphics[width=.98\textwidth]{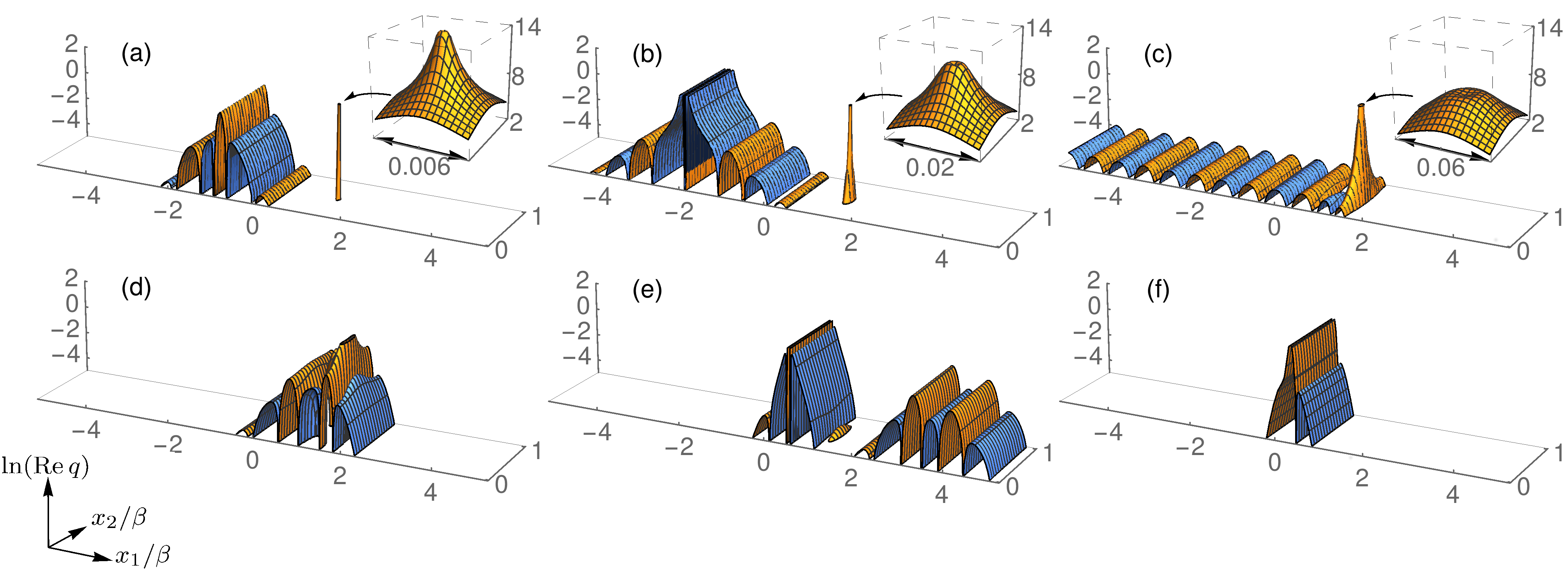}
 \caption{Logarithmic plot of $\re q(x_1,x_2)$ (cut off at $|\re q|=e^{-5}$) for the `full instanton' solution \eqref{eq:vt_wt} corresponding to Fig.\ \ref{fig:intersection}. The colors encode signs (orange: $\re q >0$, blue: $\re q <0$). In (a) and (b) ($\im \mu > 0$) the density comes in a time-dependent lump with $Q=1$ on the right and a `fractional' (almost static) lump with $Q=-\mu \beta/2\pi$ on the left. Directly at $\im \mu = 0$ ((c)), the latter becomes entirely delocalized (i.e purely oscillatory), resulting in a non-converging total charge. Note that, going from (a) to (c), the unit charged peak becomes less sharp and more extended (see insets). For $\im \mu \beta \in (-2\pi,0)$ ((d) and (e)), it splits up into two (almost static) constituents, of which only the left one remains as $\im \mu \beta < -2\pi$ (f).
 \label{fig:qre}}
\end{figure*}

\begin{figure*}[!t]
 \includegraphics[width=.98\textwidth]{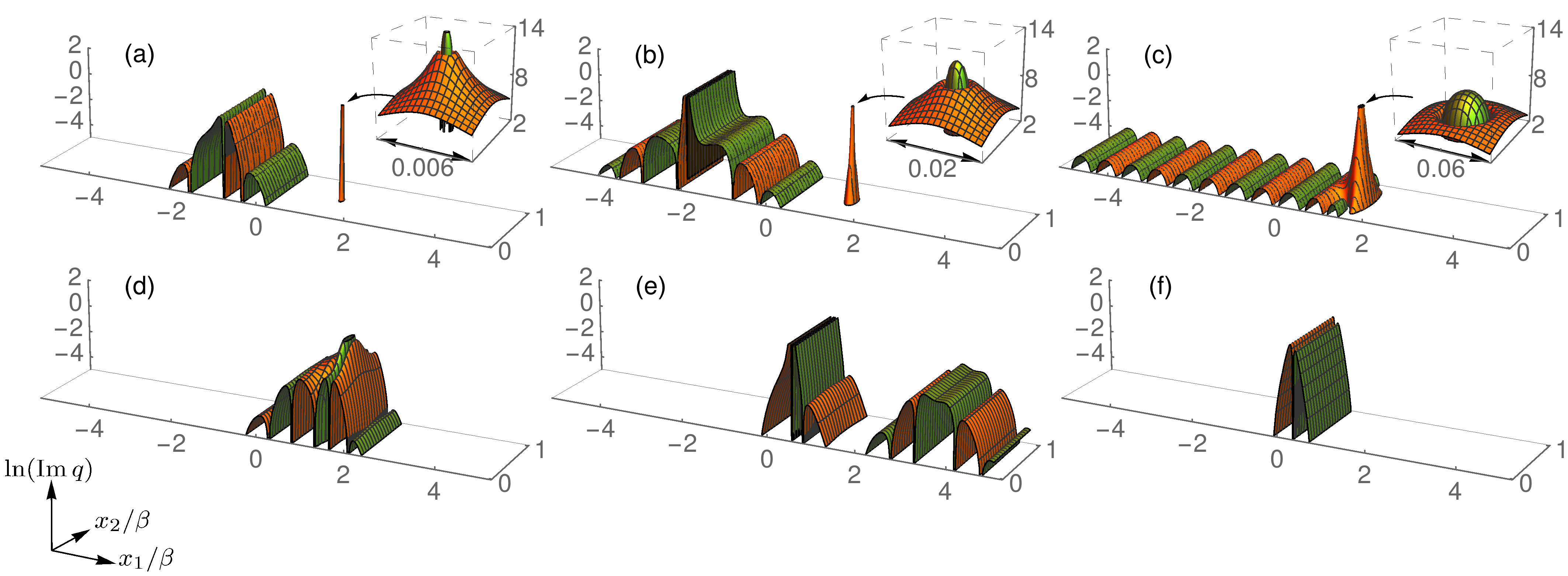}
  \caption{Logarithmic plot of $\im q(x_1,x_2)$ (orange: $\im q >0$, green: $\im q <0$, cut off at $|\im q|=e^{-5}$) corresponding to Figs.\ \ref{fig:intersection} and \ref{fig:qre}. In (a)-(c), the imaginary contribution to $Q$ essentially comes from the `fractional' constituent on the left. The sharp peaks (which are absent for purely imaginary $\mu$) at the location of the unit charge instanton (right) appear because  for $x_2/\beta \approx 1/2$ (where $\cos(2\pi x_2/\beta) = -1$), the $\mu$-dependent term in $\eqref{eq:wv}$ briefly dominates around $x_1/\beta = -\log\lambda_2/2\pi$ . However, these integrate to zero (see also the insets) and thus do not yield an imaginary contribution to the total charge. Likewise, the imaginary density of the instanton in (d) cancels and the imaginary lumps of the two constituents in (e) compensate each other, giving a total $\im Q=0$. At the transition to (f), the right one disappears at $x_1 \rightarrow \infty$, leaving a residual $\im Q = \re \mu \, \beta/2\pi$.  \label{fig:qim}}
\end{figure*}

We analyze the topological density $\Delta\log(\text{Eq.}\,\eqref{eq:wv})/4\pi$ fixing the parameters $\lambda_{1,2}$ and $\re \mu \beta= 0.8\,\pi$, while walking with the imaginary part of the chemical potential, $\im \mu \beta \in 2\pi \times[0.4, -1.2]$, through the jumps. Fig.~\ref{fig:intersection} visualizes the intersections in the logarithm, whereas real and imaginary part of the corresponding charge profiles are shown in Figs.\ \ref{fig:qre}, \ref{fig:qim}: 
(a) $\mu \beta =2\pi\,(0.4+ 0.4i)$: The charge density is split into a sharply peaked unit charged instanton (right) and a fractional constituent (left) carrying the remaining charge $Q = 0 - i\mu\beta/2\pi = 0.4 - 0.5 i$. As $\im \mu \rightarrow 0$, the latter is pushed towards $x_1=-\infty$, while the former becomes less acute and more extended (but still maintaining unit charge) (b). 
Directly at $\im \mu = 0$ (c), two zero exponents dominate as $x_1\to-\infty$, among them the one with the oscillatory part $\exp(2\, i\,\re \mu\, x_1)$. As a consequence, the density behaves asymptotically (at the corresponding end) as an inverse squared cosine, which is exactly the same non-integrability as for a single constituent discussed in Sec.~\ref{sec:example_o3}. As $\im \mu$ decreases further ((d) and (e)), the remaining unit instanton is split again into two constituents with charges proportional to $i\mu$ (left) and $i\bar{\mu}$ (right). As $\im \mu \beta \rightarrow -2\pi\; (\im \bar{\mu} \beta \rightarrow 0)$, the latter is pushed to $x_1=+\infty$ and only one lump with $Q = \mu$ survives for $\im \mu \beta < -2\pi$ ((f)), whose imaginary part is no longer compensated by the other constituent. (At $\im \mu\beta =-2\pi$, $q$ is oscillatory again, this time at $x_1 \rightarrow + \infty$.)

\section{Summary and outlook}
\label{sec:summary}

We have analyzed the equations of motion from the complex action of two-dimensional sigma models at nonzero chemical potential. By complexifying the fields we are able to push the chemical potential into properly twisted boundary conditions and to solve these equations. 

In a BPS manner, the solutions are provided by holomorphic and antiholomorphic solutions. Due to the doubled degrees of freedom, each solution is given by a pair of functions in O(3) (Eqs.~\eqref{eq_the_solutions_o3_first} and \eqref{eq_the_solutions_o3_second}) or more generally a pair of vectors 
in CP(N-1) (Eqs.~\eqref{eq_the_solutions_cpn_first} and \eqref{eq_the_solutions_cpn_second}). These become identical in the limit of purely imaginary (or vanishing) chemical potential, where the action and the saddles are real. At generic $\mu$, these objects have no meaning in isolation and thus bear certain similarities to quarks and antiquarks: Combinations of holomorphic and antiholomorphic vectors result in zero action and topological charge, like quarks and antiquarks combine into light mesons of vanishing baryon number. Combinations of just holomorphic vectors result in a nontrivial action proportional to the topological charge (on the level of densities), while combinations of antiholomorphic vectors result in a nontrivial action proportional to minus the topological charge, similar to (anti-)quarks forming heavy (anti-)baryons\footnote{Since these (anti-)baryons consist of just two (anti-)quarks, the corresponding gauge symmetry would be SU(2), 
which shows the limitation of this analogy. Another difference is that our vectors constitute just classical solutions.}. Moreover, the Fourier series of these functions are tightly linked: they must span an equal range of Fourier summands and, therefore, their moduli spaces must be of equal dimension (see the discussion at the end of Sec.~\ref{sec:comp_comp}). The existence of singularities in the topological density for specific values of the moduli parameters (as discussed for the O(3) model in Sec.~\ref{sec:example_o3}) hints at an even deeper structure within the moduli space left to future research.

One might think of looking for these objects numerically on the lattice, as was done for real twisted solutions through cooling in \cite{Brendel:2009mp}. Any method relying on minimizing the action, however, cannot be applied to the complex action. What could work is an evolution in Langevin time without noise, such that the complex drift drives the fields to the complex saddles.

It would be interesting to understand the role of these paired functions in approximate superpositions, which are needed for trans-series in (close to) neutral sectors. In some parameter space of the full instanton the solutions seem to be extremely fine-tuned giving strong action density peaks, whose imaginary parts fluctuate and cancel.

The coupling of fermions to these objects seems straightforward.
Other desirable quantities of these solutions important for physical applications are the moduli space metric and the fluctuation operator eigenmodes, the latter giving a first hint at the thimbles surrounding these saddles. For both, a certain amount of technicalities from purely imaginary $\mu$ should survive. The (anti)holomorphic formalism should also help to write down (twisted) doubly-periodic solutions for sigma models compactified in space as well. The ultimate physical goal would be to understand how the dynamical mass gap at $\mu=m$ (at low temperatures) shows up in a framework using such solutions.

\bigskip


\noindent\textbf{Acknowledgments:} FB thanks Tin Sulejmanpasic for discussions in an early stage of the study as well as Muneto Nitta and his group for discussions and their hospitality during a visit. FB also acknowledges support by a Heisenberg grant of DFG (BR 2872/6-2, BR 2872/9-1).

%

\end{document}